\documentclass[aps,prl,nofootinbib,preprintnumbers,amsmath,amssymb,latexsym,array,enumerate,letter,twocolumn,superscriptaddress]{revtex4}
\usepackage{amssymb}
\usepackage{amsmath}
\usepackage{epsfig}
\usepackage{hyperref}
\usepackage{breakurl}
\usepackage{mathtools}
\usepackage{extpfeil}
\usepackage{comment}
\usepackage{graphicx,float,color}

\makeatletter
\def\simgt{\mathrel{\lower2.5pt\vbox{\lineskip=0pt\baselineskip=0pt
           \hbox{$>$}\hbox{$\sim$}}}}
\def\simlt{\mathrel{\lower2.5pt\vbox{\lineskip=0pt\baselineskip=0pt
           \hbox{$<$}\hbox{$\sim$}}}}
\makeatother

\newcommand{\be}{\begin{eqnarray}}
\newcommand{\ee}{\end{eqnarray}}
\newcommand{\bea}{\begin{eqnarray}}
\newcommand{\eea}{\end{eqnarray}}
\newcommand{\beq}{\begin{eqnarray}}
\newcommand{\eeq}{\end{eqnarray}}

\newcommand{\sinc}{\,\textrm{sinc}\,}
\newcommand{\defS}{\mathcal{S}}

\def\lsim{\mathrel{\rlap{\lower4pt\hbox{\hskip1pt$\sim$}}
     \raise1pt\hbox{$<$}}}         
\def\gsim{\mathrel{\rlap{\lower4pt\hbox{\hskip1pt$\sim$}}
     \raise1pt\hbox{$>$}}}         

\usepackage{newfile}
\allowdisplaybreaks

\begin{document}

\title{Pulsar Polarization Arrays}

\author{Tao Liu}
\email{taoliu@ust.hk}
\affiliation{Department of Physics, The Hong Kong University of Science and Technology, Hong Kong S.A.R., P.R.China}

\author{Xuzixiang Lou}
\email{xlouaa@connect.ust.hk}
\affiliation{Department of Physics, The Hong Kong University of Science and Technology, Hong Kong S.A.R., P.R.China}

\author{Jing Ren}
\email{renjing@ihep.ac.cn}
\affiliation{Institute of High Energy Physics, Chinese Academy of Sciences, Beijing 100049, P.R.China}

\begin{abstract}

Pulsar timing arrays (PTAs) consisting of widely distributed and well-timed millisecond pulsars can serve as a galactic interferometer to measure gravitational waves. 
With the same data acquired for PTAs, we propose to develop pulsar polarization arrays (PPAs), 
to explore astrophysics and fundamental physics.
As in the case of PTAs, PPAs are best suited to reveal temporal and spatial correlations at large scales that are hard to mimic by local noise. 
To demonstrate the physical potential of PPAs, we consider detection of ultralight axion-like dark matter (ALDM), through cosmic birefringence induced by its Chern-Simons coupling. 
Because of its tiny mass, the ultralight ALDM can be generated as a Bose-Einstein condensate, characterized by a strong wave nature. 
Incorporating both temporal and spatial correlations of the signal, we show that PPAs have a potential to probe the Chern-Simons coupling up to $\sim 10^{-14}-10^{-17}$GeV$^{-1}$, with a mass range $\sim 10^{-27} - 10^{-21}$eV.

\end{abstract}

\maketitle

\section{Introduction}

Pulsars emit electromagnetic pulses with extraordinary regularity, with a period ranging from milliseconds to seconds. 
Although its mechanism is not fully clear yet, pulsars have played significant roles as astronomical clocks in testing the laws of fundamental physics.

Up to now about $3000$ pulsars have been observed. Millisecond pulsars (MSPs) formed by mass and angular momentum transfer from a companion are especially stable. Pulsar timing arrays (PTAs) consisting of many well-timed MSPs thus have been applied as a galactic interferometer to measure nanohertz gravitational waves (GWs)~\cite{Lommen:2015gbz,Tiburzi:2018txc}, complementing the  ground-based and space-based detections. 
Recently, the explorations on the PTA targets were extended to dark matter (DM) physics such as periodic oscillations in gravitational potentials induced by ultralight axion-like DM (ALDM)~\cite{Khmelnitsky:2013lxt, DeMartino:2017qsa, Porayko:2018sfa}, and Doppler or Shapiro effects induced by transiting objects associated with DM substructure~\cite{Dror:2019twh,Ramani:2020hdo,Lee:2021zqw}.

The PTA programs take up a significant amount of resources of radio telescopes. Currently, over 80 MSPs are monitored by the global PTA network
in the timespan of years. The ongoing and upcoming radio telescopes such as the 500 meter aperture spherical telescope (FAST) and the square kilometer array (SKA) are expected to significantly increase the number of well-timed MSPs to the order of 1000~\cite{Smits:2008cf, Hobbs:2014tqa, Janssen:2014dka}. 
It is apparently of great value to fully explore the physical potential of telescope resources allocated to PTAs.

Pulsars are known to be typical astrophysical sources of linearly polarized light.  The polarization profiles of their pulses are usually measured by the PTA programs for the purpose of calibration.
Thus, using the same data acquired for PTAs, we suggest to develop pulsar polarization arrays (PPAs) to detect the temporal and spatial variations of the position angle (PA). 
Currently, the polarization measurements are used to infer Faraday rotation caused by the Galactic magnetic field~\cite{Beck:2009ew, Weisberg_2010, Yan:2011gh, Yan:2011bq, NANOGrav:2021yup}, where no strong correlation between individual pulsars is expected. Nonetheless, as in the case of PTAs, PPAs are best suited to reveal temporal and spatial correlations at large scales that are hard to mimic by local noise. 

In this Letter, we demonstrate the potential reach of PPAs in testing fundamental physics by considering detection of ultralight ALDM. The ultralight ALDM 
behaves effectively as a classical scalar field in our galaxy. Its characteristic Chern-Simons interaction with photons leads to parity violation and induces cosmological birefringence~\cite{carroll1990limits, carroll1991einstein, harari1992effects}. 
Different from Faraday rotation, this ALDM-induced signal features an oscillatory signature, with the period being approximately the inverse of its mass (similar to the aforementioned oscillations in gravitational potentials~\cite{Khmelnitsky:2013lxt, DeMartino:2017qsa, Porayko:2018sfa}), and is insensitive to the light frequency.    
In the last decades, polarization measurements for a variety of astrophysical light sources have been suggested to set constraints on this coupling~\cite{carroll1990limits, Antonucci:1993sg,Ivanov:2018byi,Fujita:2018zaj,Liu:2019brz,caputo2019constraints,Chigusa:2019rra,Chen:2019fsq}. 
However, spatial correlations of the signal among individual sources have not been properly considered. 
PPAs will then play an essential role in detecting the  ALDM with its characteristic spatial correlations, 
as PTAs do for the stochastic gravitational wave backgrounds (SGWBs).

\section{ALDM-Induced Cosmic Birefringence}

An ensemble of ultralight ALDM particles forming a Bose--Einstein condensate is manifested as a classical scalar field ($a$), with its Lagrangian given by
\begin{eqnarray}
L \sim -\frac{1}{4}F_{\mu\nu}F^{\mu\nu}+ \frac{1}{2}\partial^\mu
a\partial_\mu a - \frac{1}{2}m_a^2 a^2+\frac{g}{2}a F_{\mu\nu}\tilde
F^{\mu\nu},
\end{eqnarray}
where $F^{\mu\nu}$, $\tilde{F}^{\mu\nu}$ are electromagnetic field strength and its Hodge dual, and $g$ is the Chern-Simons coupling.
While photons propagate in the ALDM field, this topological interaction corrects the dispersion relations of their positive and negative circular polarization modes in a parity-violating manner, yielding   
\begin{eqnarray}\label{dispersion}
\omega_\pm \simeq k\pm g\left(\frac{\partial a}{\partial t}+\mathbf{\nabla}
a\cdot \frac{\mathbf{k}}{k}\right)  = k \pm g\,  n^\mu \nabla_\mu a  \,.
\end{eqnarray}
Here $n^\mu$ denotes the null tangent vector to the photon path. If these photons are linearly polarized, Eq.~(\ref{dispersion}) implies a rotation of their PA, inducing the well-known effect of cosmological birefringence~\cite{carroll1990limits, carroll1991einstein}.

Pulsar light with strong linear polarizations provides a powerful tool to detect such effects~\cite{Liu:2019brz, caputo2019constraints}. Because of jitter noise induced by the stochastic variation of individual pulses~\cite{1985ApJS...59..343C,1990ApJ...349..245C}, we consider average pulse profiles to extract the ALDM-induced signal. 
With a segment of observation over the timespan $T_p$, the data for the $p$-th MSP consists of a time series of points $\Delta\theta_{p,n}\equiv\Delta\theta_p(t_n)$ for $n=1,\dots,N_p$, with each point being defined by one average profile over the folding time $\tau_\textrm{fold}$.
As the ALDM field is highly non-relativistic, its oscillating frequency $\omega_a$ is approximately $m_a$ in natural units. To preserve its oscillation imprints in the  signals, we keep $m_a\, \tau_\textrm{fold}\lesssim 1$. 
The PA rotation for the data point at $t_n$ is then given by
\begin{alignat}{1} \label{eq:Dthetapn}
\Delta\theta_{p,n} \simeq & -g\int_{C_{p,n}} ds\, n^\mu\, \nabla_\mu a(\mathbf{x},t) \\
                       =&-\frac{g}{m_a}\sum_{\mathbf{v}\in\Omega}(\Delta v)^{3/2}\alpha_\mathbf{v}\Big\{\sqrt{\rho_e f_e(\mathbf{v})}\cos\left(m_a t_n+\phi_\mathbf{v}\right)\nonumber\\
		&-\sqrt{\rho_p f_p(\mathbf{v})}\cos[m_a(t_n-L_p-\mathbf{v}\cdot\mathbf{x}_p)+\phi_\mathbf{v}]\Big\}\,.\nonumber
\end{alignat}
Here $C_{p,n}$ is the photon traveling path and $\mathbf{v}$ denotes the velocity in a latticed phase space $\Omega$. 
Because of the topological nature of the parity-violating Chern-Simons coupling, $\Delta\theta_{p,n}$ is determined by the ALDM profiles at two endpoints of this path, manifested as an ``Earth'' and a ``pulsar'' term in the second equation respectively (this highlights another difference from Faraday rotation in addition to the ones mentioned above). Accordingly, $\rho_i=\rho(\mathbf{x}_i)$ and $f_i(\mathbf{v})=f(\mathbf{x}_i,\mathbf{v})$ represent the ALDM density profile and velocity distribution near the Earth ($i=e$) and  the pulsar ($i=p$). The random amplitude $\alpha_\mathbf{v}$ and phase $\phi_\mathbf{v}$ follow the Rayleigh and uniform distributions, respectively~\cite{foster2018revealing},  encoding the stochastic nature of ALDM. We take $\mathbf{x}_e=0$ and denote by $L_p=|\mathbf{x}_p|$ the distance from the Earth to the pulsar.

For PPAs consisting of $\mathcal{N}\gg1$ pulsars, we can construct a signal vector 
\begin{eqnarray}\label{eq:signals}
	\mathbf{s}\equiv(\Delta\theta_{1,1},...,\Delta\theta_{1,N_1},..., \Delta\theta_{\mathcal{N},1},...,\Delta\theta_{\mathcal{N},N_\mathcal{N}})^T 
\end{eqnarray}
for $\Delta\theta_{p,n}$,  
with $p=1,...,\mathcal{N}$ and $n=1,...,N_p$. Integrating out the random amplitude $\alpha_\mathbf{v}$ and phase $\phi_\mathbf{v}$, the vector $\mathbf{s}$ is found to follow a multivariable Gaussian distribution with zero mean, as in the case of Ref.~\cite{foster2018revealing}.  
The statistical properties are then determined by the covariance matrix $ \mathbf{\Sigma}^{(s)}$, with $\Sigma^{(s)}_{p,n;q,m} = \langle\Delta\theta_{p,n} \Delta\theta_{q,m}\rangle$. 
To simplify the phase space integral, we assume an isotropic distribution for $\mathbf{v}$. Given that this distribution peaks sharply at the characteristic velocity of cold DM in our galaxy,  namely  $v_0\sim 10^{-3}$, 
we obtain (see supplemental material at Sec.~A for details)
\begin{eqnarray}\label{eq:Sigmaspnqm2}
\Sigma^{(s)}_{p,n;q,m}&\approx& \frac{g^2}{m_a^2}\bigg\{\rho_e\cos(m_a \Delta t) \\
&&+\sqrt{\rho_p\rho_q}\cos[m_a(\Delta t-\Delta L)]\frac{\sin y_{pq}}{y_{pq}}\nonumber\\
&&-\sqrt{\rho_e\rho_p}\cos[m_a(\Delta t-L_p)]\frac{\sin y_{ep}}{y_{ep}}\nonumber\\
&&-\sqrt{\rho_e\rho_q}\cos[m_a(\Delta t+L_q)]\frac{\sin y_{eq}}{y_{eq}}\bigg\}\,,  \nonumber 
\end{eqnarray}
with $\Delta t=t_{p,n}-t_{q,m}$, $\Delta L\equiv L_p-L_q$ and $y_{ij}=|\mathbf{x}_i-\mathbf{x}_j|/l_c$.  Here $l_c \equiv 1/(m_a v_0)$ is the de Broglie coherence length for ALDM. For pulsar-related terms, 
the distance dependence in cosines is related to the light travel time between two objects, while that for the sinc function ($\sin y_{ij}/y_{ij}$), a factor from an average over $\mathbf{v}$ direction, encodes the spatial correlation in the ALDM wave phase. 
The spatial correlation degrades when the distance $|\mathbf{x}_i-\mathbf{x}_j|$
becomes far beyond the coherence length or $y_{ij}\gg 1$. 

It is instructive to compare the ALDM-induced signal in Eq.~(\ref{eq:Sigmaspnqm2}) with the response of PTAs to GWs. 
PTAs measure the GW-induced changes of the time for pulse arrival, which can be put as Earth and pulsar terms also. 
For the isotropic SGWBs, the pulsar-related terms are suppressed in the limit of large antenna or $\omega L_p\gg1$, where the spatial correlation, manifested  as a sinc function with $y_{ij}=|\mathbf{x}_i-\mathbf{x}_j| \omega$, gets degraded quickly. The sensitivity is then dominated by the Earth-Earth term that features the Hellings-Downs quadrupolar spatial correlations among pulsars~\cite{1983ApJ...265L..39H}. As a comparison, the Earth-Earth ($\rho_e$) term in Eq.~(\ref{eq:Sigmaspnqm2}) features monopolar correlations, and cannot be distinguished from a correlated noise universal for pulsars. 
Instead, the pulsar-related terms play a more decisive role in characterizing the ALDM. Unlike the SGWBs, the ALDM has a wavenumber subject to non-relativistic suppression, manifested as the $l_c$ dependence of $y_{ij}$. The spatial correlation, arising from the wave phase, thus degrades more slowly with distance here than it does for the SGWBs, for a given wave frequency.
Because of this, the ALDM signals can get greatly enhanced for the pulsars near galactic center, where the DM halo is denser, similar to the PTA detection of the ALDM-induced oscillations in gravitational potential~\cite{DeMartino:2017qsa}. This strongly motivates the incorporation of pulsars more broadly distributed in our galaxy for both PPAs and PTAs, in particular the ones near the galactic center.

\section{Pulsar Correlations}
\label{sec:3}

To see how pulsar auto- and cross-correlations improve the ALDM detectability, let us consider a simple case where the background is dominated by white noises $\mathbf{n}$. The hypothesized data $\mathbf{d}=\mathbf{s}+\mathbf{n}$ then follow a multivariate Gaussian distribution with zero mean and a covariance matrix $\mathbf{\Sigma} = \mathbf{\Sigma}^{(s)}+\mathbf{\Sigma}^{(n)}$, where the noise contribution is $\Sigma^{(n)}_{p,n;q,m}=\lambda_p \, \delta_{p,q}\, \delta_{n,m}$, with $\lambda_p$ denoting the variance of noise. 
The likelihood function for the stochastic ALDM signal is
\begin{eqnarray}\label{eq:likelihood}
\mathcal{L}(\boldsymbol{\Theta}|\mathbf{d})=\frac{1}{\sqrt{\det[2 \pi \mathbf{\Sigma}]}} \exp \left[-\frac{1}{2} \mathbf{\tilde{d}}^{T}\cdot\mathbf{\Sigma}^{-1}\cdot \mathbf{\tilde{d}}\right] \ ,
\end{eqnarray}
where $\mathbf{\tilde{d}}$ denotes the PA data with the intrinsic effect subtracted and $\boldsymbol{\Theta}= \{g,m_a\}$ represents the model parameters (see supplemental material at Sec.~B for details).
Then given a mass $m_a$, the exclusion limit for the coupling $g$ is set by a test statistic~\cite{Cowan:2010js}
\begin{eqnarray}\label{eq:q}
q(g,m_a)\equiv2[\ln\mathcal{L}(\hat{g},m_a|\mathbf{d})-\ln\mathcal{L}(g,m_a|\mathbf{d})]\,,
\end{eqnarray}
where $\hat{g}$ maximizes $\mathcal{L}(g,m_a|\mathbf{d})$. With $q(g,m_a)=0$ for $g<\hat{g}$, the upper limit at $95\%$ confidence level $g_{95\%}$ is given by $q(g_{95\%},m_a) = 2.71$. 

The projected sensitivities are estimated using Asimov form of the test statistics, by averaging different noise realizations~\cite{Cowan:2010js}. 
For a proof-of-concept demonstration of the PPA potential, we will work in the small-signal limit in the rest of the Letter. The test statistics expanded in the second order of $\mathbf{\Sigma}^{(s)}$ is then~\cite{Foster:2020fln}  
\begin{eqnarray}\label{eq:aveq}
\langle q \rangle 
\approx \frac{1}{2}\sum_{p,q}\frac{1}{\lambda_p\lambda_q}\textrm{Tr}\left(\mathbf{\Sigma}^{(s)}_{pq}\mathbf{\Sigma}^{(s)}_{qp}\right)\, ,
\end{eqnarray}
where with a null-signal assumption we have $\hat{g}=0$. The projected $95\%$ upper limit of $g$ is set by $\langle q\rangle = 2.71$.

\begin{figure}[!h]
  \centering%
{ \includegraphics[width=7.5cm]{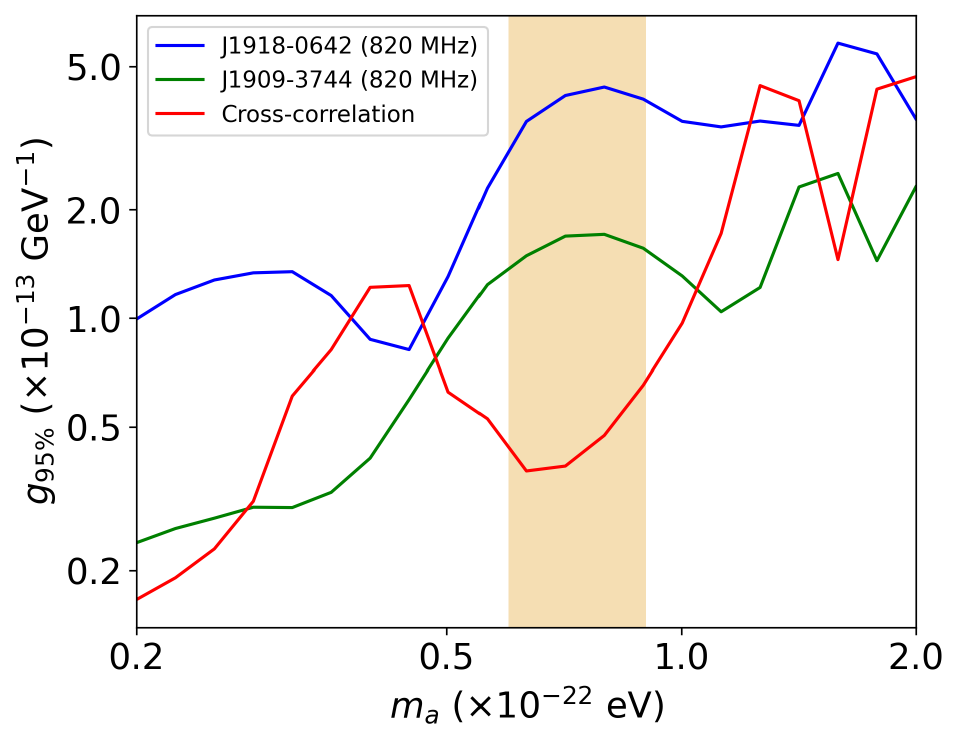}}
\caption{\label{fig:NANOGrav} 
$g_{95\%}$ obtained from the NANOGrav data of J1918-0642 (820MHz) and J1909-3744 (820MHz)~\cite{NANOGrav:2021yup}. The blue and green lines are derived from pulsar auto-correlations,  while the red line is based on their cross-correlation. }
\end{figure}

Note that the test statistics considered above is essentially a sum of the optimal filters for auto-correlation of individual pulsars ($\mathbf{\Sigma}^{(s)}_{pp}$) and  cross-correlation of different pulsars ($\mathbf{\Sigma}^{(s)}_{pq}$ with $p\neq q$) (for discussions on the optimal filters, see, e.g.,~\cite{Allen:1997ad}). 
The auto-correlation is significant when the noise variances are known well enough~\cite{Romano:2016dpx}, while the cross-correlation can distinguish a signal with long-range spatial correlations, such as the one caused by ALDM, from the noises either non-correlated or correlated in different patterns. Especially, if the backgrounds of unknown origins exist and share similar signatures of auto-correlation of the signal, the cross-correlation may become highly valuable.
As a concrete example, let us consider the sinusoidal trends of pulsar PA rotation with a period of one to two years reported by NANOGrav~\cite{NANOGrav:2021yup}. 
As shown in Fig.~\ref{fig:NANOGrav}, the auto-correlation of the data for two individual 
MSPs both indicates an anomalous peak in the khaki region, which may be explained by ALDM physics with $m_a \approx 7\times 10^{-23}$ eV. 
But, by evaluating their cross-correlation, we see null excess at this $m_a$. 
So the NANOGrav observations are very unlikely to be related to ALDM. This demonstrates the essential role played by the cross-correlation.

For the sake of completeness, we consider both auto- and cross-correlations of pulsars in the remainder of the Letter, but in a comparative manner. Below let us get a picture on their roles in determining the PPA sensitivities through their contributions to Tr$(\mathbf{\Sigma}^{(s)} \mathbf{\Sigma}^{(s)})$ in Eq.~(\ref{eq:aveq}) (see supplemental material at Sec.~B for details).

For the auto-correlation of a pulsar $p$, the matrix $\mathbf{\Sigma}^{(s)}_{pp}$ can be decomposed as: $\mathbf{\Sigma}^{(s)}_{pp}\approx A_{pp}\hat{\mathbf{\Sigma}}^{(s)}_{pp}$.
The matrix $\hat\Sigma^{(s)}_{p,n;p,m}=\cos [m_a (t_{p,n}-t_{p,m})]$ encodes temporal correlations of data points, yielding $\textrm{Tr}(\hat{\mathbf{\Sigma}}^{(s)}_{pp}\hat{\mathbf{\Sigma}}^{(s)}_{pp}) \propto N_p^2$ for the number of data points $N_p$ sufficiently large. Complementary to $\hat{\mathbf{\Sigma}}^{(s)}_{pp}$, $A_{pp}
=(g^2/m_a^2)[\rho_e+\rho_p-2\sqrt{\rho_e\rho_p}\cos(m_aL_p)\sin y_{ep}/y_{ep}]$ 
contains the messages on physical properties of axion, its halo profile and spatial distribution of pulsars. Relying on the pulsar location, the $\rho_p$ term could be either dominant over or comparable to the $\rho_e$ term in terms of their contributions to the trace. The last term of $A_{pp}$ reflects the Earth-pulsar correlation, and is subject to a suppression when $L_{p} \gg l_c$.  Approximately we have  
\begin{eqnarray}\label{eq:trsigmapp2app0}
 & &\textrm{Tr}\left(\mathbf{\Sigma}^{(s)}_{pp}\mathbf{\Sigma}^{(s)}_{pp}\right)
 \\& &\sim \frac{g^4}{m_a^4}N_p^2\Big[\rho_e+\rho_p   -2\sqrt{\rho_e\rho_p}\cos(m_aL_p)\frac{\sin y_{ep}}{y_{ep}}\Big]^2. \nonumber
\end{eqnarray}
Notably, the enhancement factor from temporal correlations, namely $N_p^2$, is not very sensitive to the distribution of sampled points and hence the data-taking strategy.

For the cross-correlation of two pulsars $p$ and $q$, the matrix $\mathbf\Sigma^{(s)}_{pq}$ can be decomposed in a similar way,  
where the temporal correlations are encoded in two matrixes: $\hat\Sigma^{(s)}_{p,n;q,m}=\cos [m_a (t_{p,n}-t_{q,m})]$ and $\hat\Sigma'^{(s)}_{p,n;q,m}=\sin [m_a (t_{p,n}-t_{q,m})]$. 
Approximately, we have
\begin{eqnarray}\label{eq:trsigmapq}
&\textrm{Tr}\left(\mathbf{\Sigma}^{(s)}_{pq}\mathbf{\Sigma}^{(s)}_{pq}\right) 
\sim \frac{g^4}{m_a^4}N_p N_q\Big[\rho_e^2  +\rho_p\rho_q\frac{\sin^2 y_{pq}}{y^2_{pq}} \nonumber \\ & +2\rho_e\sqrt{\rho_p\rho_q} \cos(m_a\Delta L)\frac{\sin y_{pq}}{y_{pq}}  + f(y_{ep}, y_{eq})\Big]\,.  
\end{eqnarray}
As in the case of auto-correlations, an enhancement factor $(\sim N_p N_q)$ appears due to temporal correlations regardless of the distribution of sampled points.  
For spatial correlation, the universal cross-correlation of the Earth terms ($\sim\rho_e^2$) always contributes. But, the most interesting contribution is from the $\rho_p\rho_q$ term which exclusively encodes the cross-correlation of physics around two pulsars, $e.g.$ $\rho_p$ and $\rho_q$,
and hence distinguishes itself from others by nature. As one of the core observables of PPAs, this term can greatly enhance the detection efficiency especially for pulsars around the galactic center and with the separation not considerably bigger than the axion coherence length, namely $\Delta x/l_c < \sqrt{ \rho_p \rho_q}/\rho_e$.

As it happens to the GW detections with PTAs, the uncertainty of pulsar distance may significantly impact the sensitivity of PPAs. In the former case, this uncertainty makes the phase of pulsar-related terms unpredictable, and the current sensitivities are dominated by the Earth term~\cite{Wang:2014ava, Zhu:2016zlx}. For PPAs, this uncertainty can cause a rapid oscillation of the Earth-pulsar terms in Eqs.~(\ref{eq:trsigmapp2app0}) and (\ref{eq:trsigmapq}) and hence eliminates their contributions.
However, there is no such oscillation for the $\rho_p\rho_q$ term in Eq.~(\ref{eq:trsigmapq}), for which the distance uncertainty enters only through the squared sinc function.
Below we will marginalize the distance uncertainty for evaluating $\langle q \rangle$.  
Currently, the pulsar distance is inferred mostly from the observed dispersion measure, where the uncertainty may reach a level $\sim 100\%$. Model-independent measurements (e.g. VLBI astrometry) can push the distance error down to $\sim 20\%$, but only for dozens of MSPs~\cite{Deller:2018zxz}.  
The situation may be greatly improved in the upcoming FAST/SKA era~\cite{lyne2012pulsar}. By using timing parallax methods, a precision $\sim20$\% could be achieved, even for MSPs around the galactic center~\cite{2011A&A...528A.108S}.

\section{Projected Sensitivities of PPAs}

Below we will demonstrate the projected sensitivity of PPAs to detect ultralight ALDM. As counterparts of PTAs at different stages, three PPA scenarios will be considered, including

\begin{itemize}
\item Near PPA (NPPA):  $100$ MSPs around the Earth, which we take as the observed MSPs in the ATNF Pulsar Catalogue~\cite{2005AJ....129.1993M} with $0.50\,\text{kpc}\leq L_p\leq 1.52\,\text{kpc}$. 
Each pulsar has $N_p=100$ data points over $T_p=10\,$yr. The noise variance is assumed to be $\lambda_p=(1\,\text{deg})^2$, typical for the current PA measurements~\cite{Yan:2011bq}.

\item Far PPA (FPPA): $100$ MSPs randomly but uniformly distributed within a $(1.0\,\text{kpc})^3$ cube around Galactic Center, $i.e$, the bulge area. $N_p$, $T_p$ and $\lambda_p$ for each pulsar are assumed the same as above.  

\item Optimal PPA (OPPA): $100$ MSPs following the distribution of MSPs in the ATNF pulsar catalog~\cite{2005AJ....129.1993M}.
As a more aggressive choice of parameters for each pulsar, we assume $T_p=30\,$yr, with an observational cadence of 1/(one week) and hence $N_p\approx 1500$ data points. $\lambda_p$ is the same as above.

\end{itemize}
These scenarios may not be fully realistic, but their performances can serve as inputs for optimizing later PTA+PPA program operations.

Another input is the galactic density profile of ultralight ALDM. 
It is known that ALDM can form a cored soliton-like structure in its halo due to quantum pressure, at a distance from the center $r\lesssim l_c$~\cite{Schive:2014dra}. But, a full simulation for the overall profile of halo with various $m_a$ values and relic-abundance shares in DM remains absent in literature. So we simply model this halo with a soliton+NFW profile~\cite{Marsh:2015xka}, namely 
\begin{equation}\label{eq:DMprofile}
  \rho(r)=  \kappa \times  \begin{cases}
   \frac{    0.019(\frac{m_a}{m_{a,0}})^{-2}(\frac{r_c}{{\rm 1 kpc}})^{-4} }{[1+9.1 \times 10^{-2} (\frac{r}{r_c})^2]^8} M_{\odot} \textrm{pc}^{-3} , & \text{for $r < l_c$}.\\
    \frac{\rho_0}{r/R_H(1+r/R_H)^2}, & \text{for $r > l_c$} \ .
  \end{cases}
\end{equation} 
which is smoothened with hyperbolic tangent function at $r\sim l_c$.
Here $r_c \approx 100 {\rm pc} \times m_{a,0}/m_a $ is obtained by fitting to the observed DM density at Galactic Center~\cite{Schive:2014dra}, with $m_{a,0} = 10^{-22}$eV.  
$\rho_0\approx 0.014\,M_\odot\text{pc}^{-3}$ and $R_H=16.1\textrm{kpc}$ are NFW parameters normalized with the local DM density near the Earth~\cite{nesti2013dark}.
$\kappa = \Omega_a/\Omega_{\rm DM}$ represents the relic abundance of ALDM, which is considerably constrained by the CMB anisotropies for $m_a\lesssim10^{-24}\,$eV~\cite{2018using}.
Constraints on $\kappa$ may arise from Lyman-$\alpha$ forest and dwarf galaxies also. These constraints however are under debate, known to suffer uncertainties from either astrophysical modeling~\cite{Ferreira:2020fam}, intrinsic properties of DM~\cite{Zhang:2017chj} or statistical treatment~\cite{Marsh:2021lqg}, and hence will not be considered.

\begin{figure}[!h]
  \centering
{ \includegraphics[width=8cm]{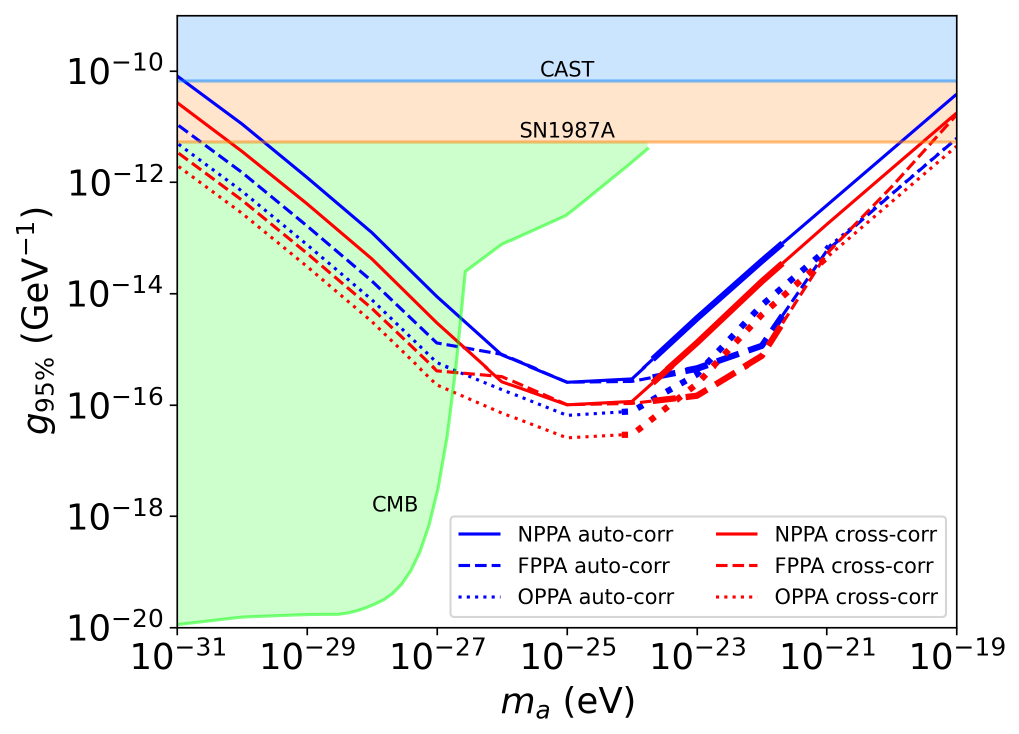}}
\caption{\label{fig:sensitivity} Projected $g_{95\%}$ with NPPA, FPPA and OPPA, based on auto- and cross-correlations of pulsars.
Segments of the curves are thickened to highlight the region of $1/T_p< m_a < N_p/T_p$, where $m_a$ can be inferred from discrete Fourier analysis.
The blue and orange regions are excluded by the CAST~\cite{collaboration2017new} and SN1987A measurements~\cite{payez2015revisiting} that are insensitive to the $m_a$ values. The green region is excluded by the CMB measurements of isotropic cosmological birefringence~\cite{refId0, 2021detection}, which is significant for $m_a \lesssim 10^{-27}$eV where ALPs behave like dark energy at the time of recombination.} 
\end{figure}

The projected $g_{95\%}$ limits as a function of $m_a$ are displayed in Fig.~\ref{fig:sensitivity}. Without loss of generality, we assume that in all PPA scenarios the temporal distribution of data points has been sampled randomly over the observation period. We present the constraints from auto- and cross-correlations separately, to highlight their respective roles in detection~\cite{footnote2}. The $\kappa$ parameter is set to its maximally allowed values by the CMB constraints~\cite{2018using} at each $m_a$ value.
The projected $g_{95\%}$ values evolve with three different stages as $m_a$ decreases: decrease first, then stay approximately flat, and finally increase. The cross-correlation limits are typically stronger than the auto-correlation ones, except in the large $m_a$ region for FPPA.

Consider NPPA first for details. For $m_a \gtrsim 10^{-24}$eV, NPPA locates in the NFW halo and hence $\rho_p \sim \rho_e$ independent of $m_a$. 
The $g_{95\%}$ value is essentially determined by the $(\rho_e + \rho_p)^2$ terms in Eq.~(\ref{eq:trsigmapp2app0}) and the Earth ($\rho_e^2$) term in Eq.~(\ref{eq:trsigmapq}). 
So we approximately have $g_{95\%} \sim m_a\, \lambda^{1/2}N_p^{-1/2}\rho_e^{-1/2}\mathcal{N}^{-1/4}$ for auto-correlation and one enhanced by $\mathcal{N}^{-1/4}$ for cross-correlation. 
The ALDM around the Earth can induce similar oscillations in CMB polarization also. But the constraint is weaker~\cite{Fedderke:2019ajk}. 
For $10^{-26} \lesssim m_a \lesssim 10^{-24}$eV, the ALDM soliton encompasses NPPA, and the $m_a$ dependence in $g^4/m_a^4$ is cancelled by that  in the soliton profile.
So $g_{95\%}$ is approximately flat except a small decrease at $m_a \sim 10^{-25}$eV due to the CMB constraints on $\kappa$~\cite{2018using}. As $m_a$ falls below $10^{-26}$eV, the Earth-pulsar terms yield a large cancellation. The $g_{95\%}$ limits get weakened as $\propto m_a^{-1}$.  Compared to NPPA, FPPA benefits from soliton-enhanced $\rho_p$. The limit is significantly improved for $m_a\sim 10^{-22}$\,eV (a benchmark model known as fuzzy DM~\cite{Khlopov:1985jw,Hu:2000ke,hui2017ultralight}) due to a large contrast of DM density $\rho_p/\rho_e\sim 10^5$.  
Above $m_a \sim 10^{-21}$eV, the cross-correlation limit degrades with decreasing coherence length $l_c$ and approaches that of NPPA.
OPPA benefits mainly from the increased $N_p$.
Given $g_{95\%} \propto \lambda^{1/2}$, the limits will be one order of magnitude stronger for all scenarios if  $\lambda \sim (0.1{\rm deg})^2$ can be achieved in the future. 
Apparently, the projected PPA limits form a great complementarity with the existing bounds.

As a final remark, we only consider white noises here, while in reality the measured PA can be contaminated by a variety of noises uncorrelated (e.g., the ones associated with individual pulsars and instruments) or with poorly understood origin~\cite{Weisberg_2010}. The auto-correlation limits in Fig.~\ref{fig:sensitivity} are likely to be deteriorated by such noises. The cross-correlation analysis, however, can distinguish the signals commonly correlating the data of pulsars, such as the ALDM one, from these noises efficiently. The generated limits are thus more robust. This is exactly the point demonstrated in Fig.~\ref{fig:NANOGrav}, and also the underlying philosophy of developing the PTAs to detect the SGWBs with characteristic spatial-correlations (see, e.g.,~\cite{1983ApJ...265L..39H}).

\section{Summary and Outlook}

In this Letter we have proposed the development of PPAs with the same data acquired for PTAs, and demonstrated their physical potential using the detection of ultralight ALDM as an example. Two important directions for future explorations can be immediately seen. 
First, we can cross-correlate PPAs and PTAs to further strengthen the ALDM detectability, given that the periodic oscillations in PA rotation have the same origin as the oscillations in gravitational potentials~\cite{Khmelnitsky:2013lxt, DeMartino:2017qsa, Porayko:2018sfa}.  
Second, PPAs provide a new tool for exploring fundamental physics. To fully resolve the physics targets such as new parity-violating origins of cosmic birefringence, it is valuable to synergize PPAs with other experimental or observational tools to improve the capability to distinguish different scenarios. 
We leave these explorations to a later work.

\begin{acknowledgments}
\subsection*{Acknowledgments}
We would greatly thank J. M. Cordes at Cornell University for highly valuable discussions on PTAs and pulsar polarization measurements, and constructive comments on this manuscript. We would also thank K. J. Lee at Peking University and the colleagues at Brown University and University of Minnesota for valuable comments on this work. T.~Liu is supported by the Collaborative Research Fund under Grant No. C6017-20G which is issued by Research Grants Council of Hong Kong S.A.R.  J. Ren is supported by the Institute of High Energy Physics, Chinese Academy of Sciences, under Contract No. Y9291220K2.
\end{acknowledgments}

\section{Supplemental Materials}

The Supplementary Materials contain additional calculation and derivation in support of the results presented in this letter. Concretely, we calculate the ALDM-induced PA rotation for linearly polarized pulsar light and its covariance matrix entries in Sec.~A, and derive the relevant test statistics and its properties in Sec.~B.

\section{A. ALDM-induced PA Rotation}
\label{sec:appA}

The ultralight ALDM field can be encoded as a superposition of particle waves, namely 
\begin{equation}
	a(\mathbf{x},t)=\frac{\sqrt{2\rho(\mathbf{x})/N_a}}{m_a}\sum_{i=1}^{N_a}\cos[m_a(t-\mathbf{v}_i\cdot\mathbf{x})+\phi_i]\,,
\end{equation}
where $\rho(\mathbf{x})$ is local DM density, $N_a$ is particle occupation number, and $\phi_i\in[0,2\pi)$ is a uniformly distributed random phase. 
In the limit of $N_a\to\infty$, this field can be approximated as
\begin{eqnarray}\label{eq:axionf}
a(\mathbf{x},t) &\approx& \frac{\sqrt{\rho(\mathbf{x})}}{m_a} \times  \\
&&  \sum_{\mathbf{v}\in\Omega}(\Delta v)^{3/2}\alpha_{\mathbf{v}}\sqrt{f_\mathbf{x}(\mathbf{v})}\cos[m_a(t-\mathbf{v}\cdot\mathbf{x})+\phi_{\mathbf{v}}]\,,  \nonumber
\end{eqnarray}
where $\Omega=\{\mathbf{v}=\Delta v(n \boldsymbol{\hat{\textbf{\i}}}+m\boldsymbol{\hat{\textbf{\j}}}+l\boldsymbol{\hat{\textbf{k}}})\,|\,n,m,l\in\mathbb{Z}\}$ denotes lattice sites in the phase space, $\Delta v=1/N_a$ is their spacing, and $f_\mathbf{x}(\mathbf{v})$ represents the DM velocity distribution at $\mathbf{x}$. Moreover, $\alpha_\mathbf{v}\in(0,+\infty)$ and $\phi_\mathbf{v}$ are parameters governed by the Rayleigh distribution $p(\alpha_\mathbf{v})=\alpha_\mathbf{v} e^{-\alpha_\mathbf{v}^2/2}$ and the uniform distribution respectively~\cite{foster2018revealing}. Their randomness reflects the stochastic nature of the  ALDM field.

Due to the parity-violation of the dispersion relations in Eq.~(\ref{dispersion}), the PA of linearly polarized pulsar light traveling across the ALDM field will be rotated by~\cite{Liu:2019brz} 
\begin{equation}
	\begin{split}
		\Delta\theta \equiv{}& \theta (t_f)-\theta (t_i)
		\simeq-g\int_{C_{p,n}}ds\,n^\mu\nabla_\mu a(\mathbf{x},t)\\
		={}&-g[a(\mathbf{x}_f,t_f)-a(\mathbf{x}_i,t_i)]\, .
	\end{split}
\end{equation}
Here $t_i$ and $t_f$ denote the moments of light emission and reception, respectively; $C_{p,n}$ is the light traveling path defined by $t_i$ and $t_f$; and $n^\mu$ represents the null tangent vector to $C_{p,n}$. For a given pulsar, $\theta (t_i)$ is determined by the intrinsic PA of its light, {\it i.e.}, $\theta (t_i) = \theta_{\rm in}$. The value of $\theta_{\rm in}$ is unknown and usually assumed to be a constant. Notably, $\Delta\theta$ relies on the amplitude of ALDM field at the light emission and reception points only.

For the $p$-th pulsar in the array, the PA rotation for one pulse received at $t_n$ is then given by 
\begin{eqnarray}\label{eq:Dthetapn2}
\Delta\theta_{p,n} &=& \theta_{p}(t_n)-\theta_{{\rm in},p} \\
&\simeq& -g[a(\mathbf{x}_e,t_n)-a(\mathbf{x}_p,t_n-L_p)] \nonumber\\
                       &=&-\frac{g}{m_a}\sum_{\mathbf{v}\in\Omega}(\Delta v)^{3/2}\alpha_\mathbf{v}\Big\{\sqrt{\rho_e f_e(\mathbf{v})}\cos\left(m_a t_n+\phi_\mathbf{v}\right)\nonumber\\
		&&-\sqrt{\rho_p f_p(\mathbf{v})}\cos[m_a(t_n-L_p-\mathbf{v}\cdot\mathbf{x}_p)+\phi_\mathbf{v}]\Big\}\,.\nonumber
\end{eqnarray}
Here we use a simplified notation $\rho_i\equiv\rho(\mathbf{x}_i)$, $f_i(\mathbf{v})\equiv f_{\mathbf{x}_i}(\mathbf{v})$, with $i=p, e$. We take $\mathbf{x}_e=0$, such that  the distance from the Earth to the pulsar $L_p=|\mathbf{x}_p|$. 
When the folding time $\tau_\textrm{fold} < 1/m_a$, Eq.~(\ref{eq:Dthetapn2}) can be approximately applied to the average profile for each observational point, yielding Eq.~(\ref{eq:Dthetapn}) in the main text. 

The signal vector $\mathbf{s}$ (see Eq.~(\ref{eq:signals})) follows a multivariable Gaussian distribution, with a mean of zero over the random phase $\phi_\mathbf{v}$. 
Its statistical properties are described by a covariance matrix $\mathbf{\Sigma}^{(s)}$,  
$i.e.$, the two-point correlation function of $\Delta\theta_{p,n}$, with the entries being given by    
\begin{widetext}
\begin{eqnarray}\label{eq:Sigmaelement1}
		\Sigma_{p,n;q,m}^{(s)}&=&\langle\Delta\theta_{p,n}\Delta\theta_{q,m}\rangle  \nonumber  \\
		&=&\frac{g^2}{m_a^2}\sum_{\mathbf{v}\in\Omega}\sum_{\mathbf{v}'\in\Omega}(\Delta v)^3\langle\alpha_\mathbf{v}\alpha_{\mathbf{v}'}\rangle
		\Big\{\rho_ef_e(\mathbf{v})\,\langle\cos(m_at_n+\phi_\mathbf{v})\cos(m_at_m+\phi_{\mathbf{v}'})\rangle   \nonumber  \\ 
		&&	
		+\sqrt{\rho_p\rho_qf_p(\mathbf{v})f_q(\mathbf{v}')}\,\langle\cos[m_a(t_n-L_p-\mathbf{v}\cdot\mathbf{x}_p)+\phi_\mathbf{v}]
		\cos[m_a(t_m-L_q-\mathbf{v}'\cdot\mathbf{x}_q)+\phi_{\mathbf{v}'}]\rangle \nonumber  \\ 
		&&	 
		 -\sqrt{\rho_p\rho_ef_p(\mathbf{v})f_e(\mathbf{v}')}
	        \langle\cos[m_a(t_n-L_p-\mathbf{v}\cdot\mathbf{x}_p)+\phi_\mathbf{v}]\cos(m_at_m+\phi_{\mathbf{v}'})\rangle  \nonumber  \\ 
		&&
		-\sqrt{\rho_e\rho_qf_e(\mathbf{v})f_q(\mathbf{v}')}\,\langle\cos(m_at_n+\phi_\mathbf{v})
		\cos[m_a(t_m-L_q-\mathbf{v}'\cdot\mathbf{x}_q)+\phi_{\mathbf{v}'}]\rangle \Big\} \nonumber  \\
		&\approx&  \frac{g^2}{m_a^2}\int d^3\mathbf{v}\,\Big\{\rho_ef_e(\mathbf{v})\cos(m_a\Delta t)
		+\sqrt{\rho_p\rho_qf_p(\mathbf{v})f_q(\mathbf{v})}\cos[m_a(\Delta t-\Delta L-\mathbf{v}\cdot\Delta\mathbf{x})] \nonumber  \\ 
		&&
		-\sqrt{\rho_p\rho_ef_p(\mathbf{v})f_e(\mathbf{v})}\cos[m_a(\Delta t-L_p-\mathbf{v}\cdot\mathbf{x}_p)]
		-\sqrt{\rho_e\rho_qf_e(\mathbf{v})f_q(\mathbf{v})}\cos[m_a(\Delta t+L_q+\mathbf{v}\cdot\mathbf{x}_q)] \Big\}\, . 
\end{eqnarray}
\end{widetext}
Here $\Delta t=t_n-t_m$, $\Delta L=L_p-L_q$, $\Delta\mathbf{x}=\mathbf{x}_p-\mathbf{x}_q$. In the last step, $ \langle\cos(a+\phi_\mathbf{v})\cos(b+\phi_{\mathbf{v}'})\rangle=\frac{1}{2}\cos(a-b)\delta_{\mathbf{v},\mathbf{v}'}$ and $\langle\alpha_\mathbf{v}^2\rangle=2$ have been applied. The summation is converted to an integral by taking the limit of $\Delta v\to0$.

To take this integral over the phase space, we further assume that the ALDM velocity distributes isotropically and peak sharply at the virial velocity $v_0$ of our galaxy: $f_\mathbf{x}(\mathbf{v})\approx (4\pi v^2)^{-1}\delta(v-v_0)$ with $v_0\sim 10^{-3}$. We then have Eq.~(\ref{eq:Sigmaspnqm2}) in the main text
\begin{eqnarray}\label{eq:Sigmaspnqm2p}
\Sigma^{(s)}_{p,n;q,m}&\approx& \frac{g^2}{m_a^2}\bigg\{\rho_e\cos(m_a \Delta t) \\
&&+\sqrt{\rho_p\rho_q}\cos[m_a(\Delta t-\Delta L)]\frac{\sin y_{pq}}{y_{pq}}\nonumber\\
&&-\sqrt{\rho_e\rho_p}\cos[m_a(\Delta t-L_p)]\frac{\sin y_{ep}}{y_{ep}}\nonumber\\
&&-\sqrt{\rho_e\rho_q}\cos[m_a(\Delta t+L_q)]\frac{\sin y_{eq}}{y_{eq}}\bigg\}\,,  \nonumber 
\end{eqnarray}
where $y_{ij}=|\mathbf{x}_i-\mathbf{x}_j|/l_c$, with $l_c \equiv 1/(m_a v_0)$ being the de Broglie coherence length of ALDM. 
The sinc function $\sinc y_{ij}\equiv \sin y_{ij}/y_{ij}$ arises from the integral $\frac{1}{2}\int_0^\pi\cos(a+b\cos\theta)\sin\theta d\theta=\cos a \sin b/b$, which reflects the spatial correlation in the ALDM wave phase.

\section{B. Likelihood Analysis}

To analyze the ALDM-induced PA rotation, we adopt a likelihood function for stochastic signal:
\begin{eqnarray}\label{eq:likelihood2}
\mathcal{L}'(\boldsymbol{\Theta}'|\mathbf{d})&=&\frac{1}{\sqrt{\det[2 \pi \mathbf{\Sigma}]}}  \times  \\
&& \exp \left[-\frac{1}{2}\left(\mathbf{d}-\boldsymbol{\theta}_{\rm in}\right)^{T}\mathbf{\Sigma}^{-1}\left(\mathbf{d}-\boldsymbol{\theta}_{\rm in}\right)\right]   \ . \nonumber
\end{eqnarray}
Here, $\mathbf{d}$ denote the PA data, with a dimension $N_d = \sum_{p=1}^{\mathcal{N}} N_p$. 
$\boldsymbol{\Theta}'=\{g, m_a, \boldsymbol{\theta}_{\rm in}\}$ represent the ALDM and PPA parameters, with $\boldsymbol{\theta}_{\rm in}=(\theta_{{\rm in}, 1}, ..., \theta_{{\rm in}, \mathcal{N}})$ being the intrinsic PAs for individual pulsars. 
As their values are unknown and usually assumed to be stationary for a given pulsar,  $\boldsymbol{\theta}_{\rm in}$ are essentially a set of nuisance parameters in the analysis. 
$\mathbf{\Sigma} = \mathbf{\Sigma}^{(s)}+\mathbf{B}$ is an $N_d\times N_d$ covariance matrix for multivariate Gaussian distribution, where $ \mathbf{\Sigma}^{(s)}$ and $\mathbf{B}$ denote the contributions from signal (see Eq.~(\ref{eq:Sigmaspnqm2p})) and noise, respectively.

For a given mass $m_a$, the test statistics for setting an upper limit on the coupling $g$ is given by~\cite{Cowan:2010js}
\begin{eqnarray}\label{eq:q0}
\small
q(g,m_a)\equiv2\left[\ln\mathcal{L}'\left(\hat{g}, \boldsymbol{\hat\theta}_{\rm in},m_a|\mathbf{d}\right)-\ln\mathcal{L}'\left(g,\boldsymbol{\hat{\hat\theta}}_{\rm in}, m_a|\mathbf{d}\right)\right]\,,\nonumber\\
\end{eqnarray}
where $\boldsymbol{\hat{\hat\theta}}_{\rm in}$ is the conditional maximum-likelihood estimator,  and $\hat{g}$, $\boldsymbol{\hat\theta}_{\rm in}$ are the maximum-likelihood estimators with $\boldsymbol{\hat\theta}_{\rm in}= \boldsymbol{\hat{\hat\theta}}_{\rm in}(\hat{g})$. $\boldsymbol{\hat{\hat\theta}}_{\rm in}$ can be derived from $\partial \ln\mathcal{L}'/\partial \boldsymbol{\theta}_{\rm in}|_{\boldsymbol{\hat{\hat\theta}}_{\rm in}}=0$.  This immediately gives 
\begin{eqnarray}\label{eq:hatthetain}
\boldsymbol{\hat{\hat\theta}}_{\rm in}=\mathbf{d}^T \,\mathbf{\Sigma}^{-1}\,\mathbb{I}\; \mathcal{M}^{-1}\,,
\end{eqnarray} 
with
\begin{eqnarray}\label{eq:matrixIM}
\mathbb{I}&\equiv& \left(
\begin{array}{cccc}
I_1& 0 & ... & 0 \\
0 & I_2 &... & 0 \\
... & ... & ... & ... \\
0 & 0 & ... & I_\mathcal{N}
\end{array}
\right)_{N_d\times\mathcal{N}},\quad 
I_p=(1,1,...,1)_{1\times N_p}^T\nonumber\\
\mathcal{M}&\equiv& \left( 
\begin{array}{ccc}
I_1^T (\mathbf{\Sigma}^{-1})_{11}I_1& ... & I_1^T (\mathbf{\Sigma}^{-1})_{1\mathcal{N}}I_\mathcal{N}\\
... & ... & ... \\
I_\mathcal{N}^T (\mathbf{\Sigma}^{-1})_{\mathcal{N}1}I_1 & ... &  I_\mathcal{N}^T (\mathbf{\Sigma}^{-1})_{\mathcal{N}\mathcal{N}}I_\mathcal{N}
\end{array}
\right)_{\mathcal{N}\times \mathcal{N}} \ .
\end{eqnarray}
Here,  $\mathbb{I}$ is block-diagonal and constructed by the vector $I_p$ of dimension $N_p$, and $\mathcal{M}$ is constructed by the corresponding blocks of $\mathbf{\Sigma}^{-1}$.
Substituting $\boldsymbol{\hat{\hat\theta}}_{\rm in}$ in Eq.~(\ref{eq:q0}) with Eq.~(\ref{eq:hatthetain}), one can find 
\begin{eqnarray}
q(g,m_a)=2 [\ln\mathcal{L}(\hat{g},m_a|\mathbf{d})-\ln\mathcal{L}(g,m_a|\mathbf{d})]\,,
\end{eqnarray}
with the profile likelihood 
\begin{eqnarray}\label{eq:likelihood3}
\mathcal{L}(\boldsymbol{\Theta}|\mathbf{d})
&=& \mathcal{L}'(\boldsymbol{\Theta}, \boldsymbol{\hat{\hat\theta}}_{\rm in} |\mathbf{d})\nonumber\\
&=&\frac{1}{\sqrt{\det[2 \pi \mathbf{\Sigma}]}} \exp \left[-\frac{1}{2}\mathbf{\tilde{d}}^{T}\mathbf{\Sigma}^{-1}\mathbf{\tilde{d}}\right]\,,
\end{eqnarray}
where $\boldsymbol{\Theta}=\{g, m_a\}$ and $\mathbf{\tilde{d}}=\mathbf{d}-\boldsymbol{\hat{\hat\theta}}_{\rm in}$. They are exactly the likelihood function in Eq.~(\ref{eq:likelihood}) and the test statistics in Eq.~(\ref{eq:q}).

The Asimov form of the test statistics provides a convenient way to estimate the projected sensitivities. The exclusion limit in this scheme can be derived by requiring the average of the test statistics over all possible noise realizations, denoted as $\langle q \rangle$, to be bigger than certain statistical threshold~\cite{Cowan:2010js}. Under the assumption of $\hat{g}=0$, we then have 
\begin{eqnarray}\label{eq:qfull}
\langle q \rangle&=& \textrm{Tr}\left[\left\langle\left(\mathbf{d}-\boldsymbol{\hat{\hat\theta}}_{\rm in}\right)\left(\mathbf{d}-\boldsymbol{\hat{\hat\theta}}_{\rm in}\right)^T\right\rangle \mathbf{\Sigma}^{-1}\right]\nonumber\\
&&-\textrm{Tr}\left[\left\langle\left(\mathbf{d}-\hat{\boldsymbol{\theta}}^{\rm in}\right)\left(\mathbf{d}-\hat{\boldsymbol{\theta}}^{\rm in}\right)^T\right\rangle \mathbf{B}^{-1}\right]\nonumber\\
&&+\ln\det \mathbf{\Sigma}/\ln\det \mathbf{B}\nonumber\\
&=&\langle q \rangle_0+\langle q \rangle_{\rm in}\,,
\end{eqnarray}
with 
\begin{eqnarray}
\langle q \rangle_0&=&\textrm{Tr}\left[\mathbf{B}(\mathbf{\Sigma}^{-1}-\mathbf{B}^{-1})\right]+\ln\det \mathbf{\Sigma}/\ln\det \mathbf{B}\,,\nonumber\\
\langle q \rangle_{\rm in}&=&\textrm{Tr}\left[(\mathbf{\Sigma}-\mathbf{B})\mathbf{\Sigma}^{-1}\mathbb{I}\,\mathcal{M}^{-1}\mathbb{I}^T\mathbf{\Sigma}^{-1}\right]\, .
\end{eqnarray}
Here $\langle q \rangle_0$ denotes the contributions irrelevant to the nuisance parameters $\boldsymbol{\theta}_{\rm in}$,
while $\langle q \rangle_{\rm in}$ denotes the non-trivial corrections from $\boldsymbol{\theta}_{\rm in}$.

In the limit of small signal, i.e. small $\mathbf{\Sigma}^{(s)} \mathbf{B}^{-1}$, $\langle q \rangle_0$ and $\langle q \rangle_{\rm in}$ are given by
\begin{eqnarray}
\langle q \rangle_0&\approx&\frac{1}{2}\textrm{Tr}\left[\mathbf{\Sigma}^{(s)}\mathbf{B}^{-1}\mathbf{\Sigma}^{(s)}\mathbf{B}^{-1}\right]  \ ,  \nonumber\\
\langle q \rangle_{\rm in}&\approx&\textrm{Tr}\left[\mathbf{\Sigma}^{(s)}\mathbf{B}^{-1}\mathbb{I}\,\mathcal{M}^{-1}\mathbb{I}^T\mathbf{B}^{-1}\right] 
\end{eqnarray}
at the leading order. By further assuming white noise only, {\it i.e.}, $B_{p,n;q,m}=\lambda_p \, \delta_{p,q}\, \delta_{n,m}$, we have 
\begin{eqnarray} \label{eq:qs}
\langle q \rangle_0&=& \langle q \rangle_{0, \rm auto} + \langle q \rangle_{0, \rm cross}  \ ,  \nonumber\\
\langle q \rangle_{0, \rm auto} &\approx& \frac{1}{2} \sum_p \frac{1}{\lambda_p^2}\textrm{Tr}\left[\mathbf{\Sigma}^{(s)}_{pp}\mathbf{\Sigma}^{(s)}_{pp}\right]  \ , \nonumber\\
\langle q \rangle_{0, \rm cross} &\approx&    \frac{1}{2} \sum_{p\neq q} \frac{1}{\lambda_p\lambda_q}\textrm{Tr}\left[\mathbf{\Sigma}^{(s)}_{pq}\mathbf{\Sigma}^{(s)}_{qp}\right]  \ , \nonumber\\
\langle q \rangle_{\rm in}&\approx&\sum_p\frac{1}{N_p \lambda_p} I_p^T\mathbf{\Sigma}^{(s)}_{pp}I_p \,.
\end{eqnarray}
Interestingly, $\langle q \rangle_{0}$ is exactly the sum of the optimal filters for auto-correlation ($\langle q \rangle_{0, \rm auto}$) and cross-correlation ($\langle q \rangle_{0, \rm cross}$) (for discussions on the optimal filters, see, e.g.,~\cite{Allen:1997ad}). The intrinsic PA related contribution $\langle q \rangle_{\rm in}$ is linearly dependent on $\mathbf{\Sigma}^{(s)}_{pp}$. It contributes to the auto-correlation test statistics, but mildly in a usual case. 
For simplicity, we will not consider $\langle q \rangle_{\rm in}$ in this proof-of-concept work. Then with $\langle q \rangle\approx \langle q \rangle_{0}$ we have Eq.~(\ref{eq:aveq}).

For auto-correlation of the $p$-th pulsar, the matrix $\mathbf{\Sigma}^{(s)}_{pp}$ can be decomposed as
\begin{eqnarray}
\mathbf{\Sigma}^{(s)}_{pp}\approx A_{pp}\hat{\mathbf{\Sigma}}^{(s)}_{pp}\,,
\end{eqnarray}
where the matrix $\hat\Sigma^{(s)}_{pp,nm}=\cos [m_a (t_{p,n}-t_{p,m})]$ encodes the effect of temporal correlation, and the coefficient 
$A_{pp}=(g^2/m_a^2)[\rho_e+\rho_p-2\sqrt{\rho_e\rho_p}\cos(m_aL_p)\sinc y_{ep}]$ contains the messages on the axion properties, the ALDM halo profile and the pulsar spatial distribution. In an ideal case where the $N_p$ data points of this pulsar are sampled uniformly (or with a constant separation $\tau_p$) for an observation period $T_p=N_p\tau_p$, we have 
\begin{eqnarray}\label{eq:sigmapphat}
 \textrm{Tr}\left(\hat{\mathbf{\Sigma}}^{(s)}_{pp}\hat{\mathbf{\Sigma}}^{(s)}_{pp}\right)  
=\frac{1}{2}N_p^2\Big[1+\defS^2(m_a T_p)\Big]  \ , 
\end{eqnarray}
where  
\begin{eqnarray}
\defS(y)=\frac{\sin y}{N_p \sin (y/N_p)}\,.
\end{eqnarray}
For $m_a \ll 1/T_p$, we have $\textrm{Tr}\left(\hat{\mathbf{\Sigma}}^{(s)}_{pp}\hat{\mathbf{\Sigma}}^{(s)}_{pp}\right)  \sim N_p^2$.  
When $1/T_p\ll m_a\ll 1/\tau_p$, we have $\defS(y)\sim \sinc y$ and $\textrm{Tr}\left(\hat{\mathbf{\Sigma}}^{(s)}_{pp}\hat{\mathbf{\Sigma}}^{(s)}_{pp}\right) \to N_p^2/2$. 
If $1/\tau_p\ll m_a\ll 1/\tau_\textrm{fold}$,
$\defS(y)$ oscillates between $1$ and $-1$.   
$\textrm{Tr}\left(\hat{\mathbf{\Sigma}}^{(s)}_{pp}\hat{\mathbf{\Sigma}}^{(s)}_{pp}\right)$ 
is thus proportional to $N_p^2$ with the magnitude of its coefficient $\lesssim 1$.

\begin{figure}[!h]
  \centering
{ \includegraphics[width=6.5cm]{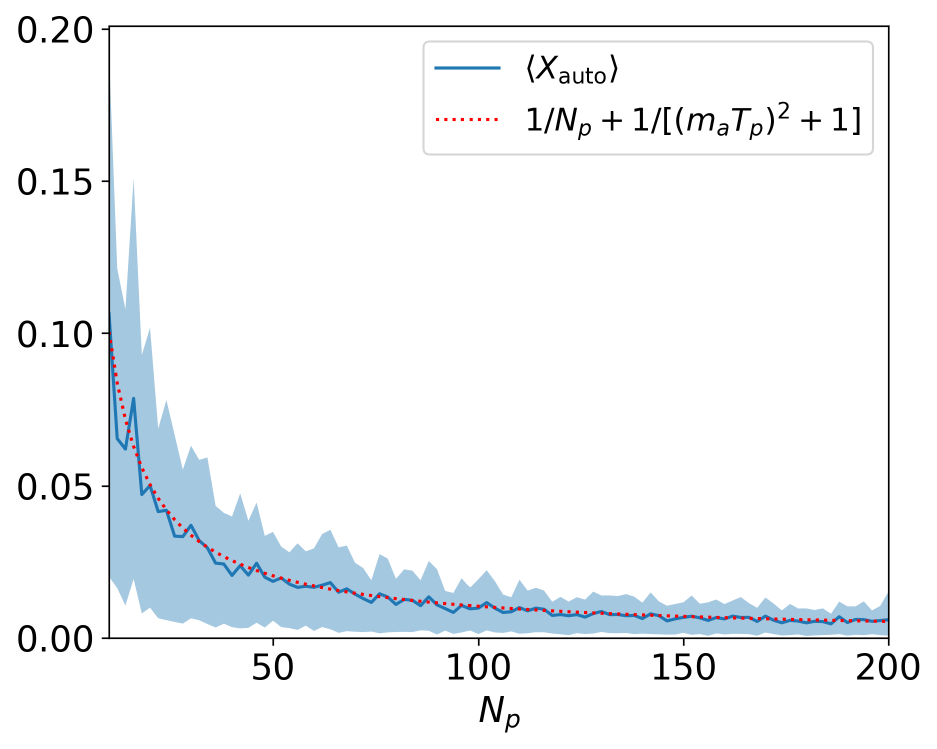}}
{ \includegraphics[width=6.5cm]{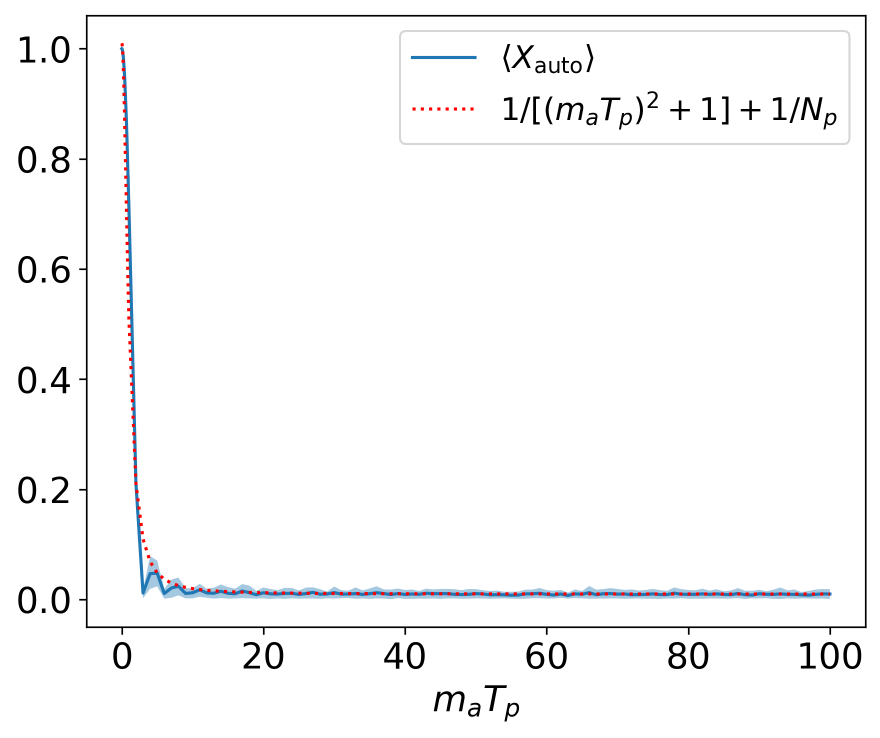}}
\caption{\label{fig:trpp} 
Dependence of $\langle X_{\rm auto}\rangle$ (solid) on $N_p$ for $m_a T_p=48$ (top) and on $m_a T_p$ for $N_p=100$ (bottom). Here $\langle X_{\rm auto}\rangle$ is taken as a mean over 100 random samplings of $\{t_{p,n}; n = 1, ......, N_p\}$. The shaded band denotes its variance at 68\% C.L. The red dotted line in each panel represents a simple fit to the trend of $\langle X_{\rm auto}\rangle$. $m_a T_p=48$ implies $T_p=10$~yr for $m_a=10^{-22}$~eV.}
\end{figure}

For a more general treatment where the data points are taken non-uniformly in time sequence, we introduce a counterpart variable of $\defS^2(m_a T_p)$ (see Eq.~(\ref{eq:sigmapphat})) 
\begin{eqnarray}
X_{\rm auto} \equiv \frac{\textrm{Tr}(\hat{\mathbf{\Sigma}}^{(s)}_{pp}\hat{\mathbf{\Sigma}}^{(s)}_{pp}) }{(N_p^2/2)}-1\,.
\end{eqnarray}
Then, we generate 100 random samplings of $\{t_{p,n}\in[0, T_p]; n = 1, ......, N_p\}$, and demonstrate in Fig.~\ref{fig:trpp} how the $X_{\rm auto}$ distribution over these samplings varies with $N_p$ and $m_a T_p$. 
As shown in this figure, the mean $\langle X_{\rm auto}\rangle$ drops quickly to zero as $N_p$ and $m_a T_p$ increase, with a variance of $\lesssim 0.1$. 
This trend can be well described as 
\begin{eqnarray}
\langle X_{\rm auto}\rangle \rightarrow \frac{1}{N_p}+\frac{1}{(m_a T_p)^2 +1} \ . 
\end{eqnarray}
As $N_p \gg 1$ usually, we have $X_{\rm auto} \lesssim 1$ within the statistical range of 68\%. 
The $1/N_p$ term here never contributes to $\textrm{Tr}(\hat{\mathbf{\Sigma}}^{(s)}_{pp}\hat{\mathbf{\Sigma}}^{(s)}_{pp})$ significantly. 
Thus we have the approximate relation in Eq.~(\ref{eq:trsigmapp2app0})
\begin{eqnarray}
\textrm{Tr}\left(\mathbf{\Sigma}^{(s)}_{pp}\mathbf{\Sigma}^{(s)}_{pp}\right) \sim N_p^2 A_{pp}^2   \ .
\end{eqnarray}

For the cross-correlation of two pulsars $p$ and $q$, the matrix $\mathbf\Sigma^{(s)}_{pq}$ can be decomposed as 
\begin{eqnarray}
\mathbf\Sigma^{(s)}_{pq}\approx A_{pq}\,\hat{\mathbf{\Sigma}}^{(s)}_{pq}+ A'_{pq}\,\hat{\mathbf{\Sigma}}'^{(s)}_{pq}\,. 
\end{eqnarray}
The temporal correlations are now encoded in two matrixes: $\hat\Sigma^{(s)}_{pq,nm}=\cos [m_a (t_{p,n}-t_{q,m})]$ and $\hat\Sigma'^{(s)}_{pq,nm}=\sin [m_a (t_{p,n}-t_{q,m})]$. The coefficients $A_{pq}$ and $A'_{pq}$ are given by 
\begin{eqnarray}\label{eq:Apq}
A_{pq}&=&\frac{g^2}{m_a^2}\big[\rho_e+\sqrt{\rho_p\rho_q}\cos(m_a \Delta L)\sinc y_{pq}-\sqrt{\rho_e\rho_p}  \times \nonumber\\
&&\cos(m_aL_p)\sinc y_{ep}-\sqrt{\rho_e\rho_q}\cos(m_aL_q)\sinc y_{eq}\big]   \ , \nonumber\\
A'_{pq}&=&\frac{g^2}{m_a^2}\big[\sqrt{\rho_p\rho_q}\sin(m_a \Delta L)\sinc y_{pq}
-\sqrt{\rho_e\rho_p} \times \\ 
&& \sin(m_aL_p) \sinc y_{ep}+\sqrt{\rho_e\rho_q}\sin(m_aL_q)\sinc y_{eq}\big]\,. \nonumber 
\end{eqnarray}

Again let us consider first the ideal case of uniform sampling. Assuming that the time series of these two pulsars contain $N_p$ and $N_q$ data points with a temporal separation $\tau_p$ and $\tau_q$ respectively, we find  
\begin{alignat}{1}
\label{eq:sigmapqhat}
\textrm{Tr}\left(\hat{\mathbf{\Sigma}}^{(s)}_{pq}\hat{\mathbf{\Sigma}}^{(s)}_{qp}\right)
&=\frac{1}{2}N_p N_q\big[1+\defS(m_a T_p)\defS(m_a T_q)\cos\Phi\big]  \,  ,   \nonumber\\
\textrm{Tr}\left(\hat{\mathbf{\Sigma}}'^{(s)}_{pq}\hat{\mathbf{\Sigma}}'^{(s)}_{qp}\right)
&=\frac{1}{2}N_p N_q\big[1-\defS(m_a T_p)\defS(m_a T_q)\cos\Phi\big]  \,  ,  \nonumber\\
\textrm{Tr}\left(\hat{\mathbf{\Sigma}}^{(s)}_{pq}\hat{\mathbf{\Sigma}}'^{(s)}_{qp}\right)
&=\frac{1}{2}N_p N_q\defS(m_a T_p)\defS(m_a T_q)\sin\Phi   \, ,
\end{alignat}
where $\Phi=m_a(t_{p,N_p}-t_{q,N_q}+t_{p,1}-t_{q,1})$ measures the time difference of the initial and final data points between the two pulsars. 
Given that the value of $\defS(y)$ varies between $[-1, 1]$, the first two and the last traces are approximately proportional to $N_pN_q$, with a coefficient $\in [0, 1]$ and $\in [-\frac{1}{2},\frac{1}{2}]$, respectively.

For the more general case where the data points are non-uniformly taken in time sequence, Eq.~(\ref{eq:sigmapqhat}) becomes
\begin{eqnarray}\label{eq:sigmapqhat_2}
&& \textrm{Tr}\left(\hat{\mathbf{\Sigma}}^{(s)}_{pq}\hat{\mathbf{\Sigma}}^{(s)}_{qp}\right)  \ = \  \\
&& \quad  \frac{N_pN_q}{2}  
 \Bigg [1+  \frac{1}{N_pN_q}\sum_{n=1}^{N_p}\sum_{m=1}^{N_q} \cos[2m_a(t_{q,m}-t_{p,n})] \Bigg]  \ ,  \nonumber\\
&&\textrm{Tr}\left(\hat{\mathbf{\Sigma}}'^{(s)}_{pq}\hat{\mathbf{\Sigma}}'^{(s)}_{qp}\right) \ = \  \nonumber \\
&& \quad  \frac{N_pN_q}{2}  
 \Bigg [1-  \frac{1}{N_pN_q}\sum_{n=1}^{N_p}\sum_{m=1}^{N_q} \cos[2m_a(t_{q,m}-t_{p,n})] \Bigg] \ , \nonumber
\end{eqnarray}
and 
\begin{eqnarray}
\textrm{Tr}\left(\hat{\mathbf{\Sigma}}^{(s)}_{pq}\hat{\mathbf{\Sigma}}'^{(s)}_{qp}\right) = 
\frac{1}{2}\sum_{n=1}^{N_p}\sum_{m=1}^{N_q} \sin[2m_a(t_{q,m}-t_{p,n})] \, .
\end{eqnarray}
Similar to the case of auto-correlation, we introduce two intermediate variables 
\begin{eqnarray}\label{eq:sigmapqhat_3}
X_{\rm cross} &=& \frac{\textrm{Tr}\left(\hat{\mathbf{\Sigma}}^{(s)}_{pq}\hat{\mathbf{\Sigma}}^{(s)}_{qp}\right)}{N_p N_q/2}-1
=-\frac{\textrm{Tr}\left(\hat{\mathbf{\Sigma}}'^{(s)}_{pq}\hat{\mathbf{\Sigma}}'^{(s)}_{qp}\right)}{N_p N_q/2}+1 \ , \nonumber   \\
X'_{\rm cross} &=& \frac{\textrm{Tr}\left(\hat{\mathbf{\Sigma}}^{(s)}_{pq}\hat{\mathbf{\Sigma}}'^{(s)}_{qp}\right)}{N_pN_q/2} \ .
\end{eqnarray}

\begin{figure} 
  \centering
{ \includegraphics[width=6.5cm]{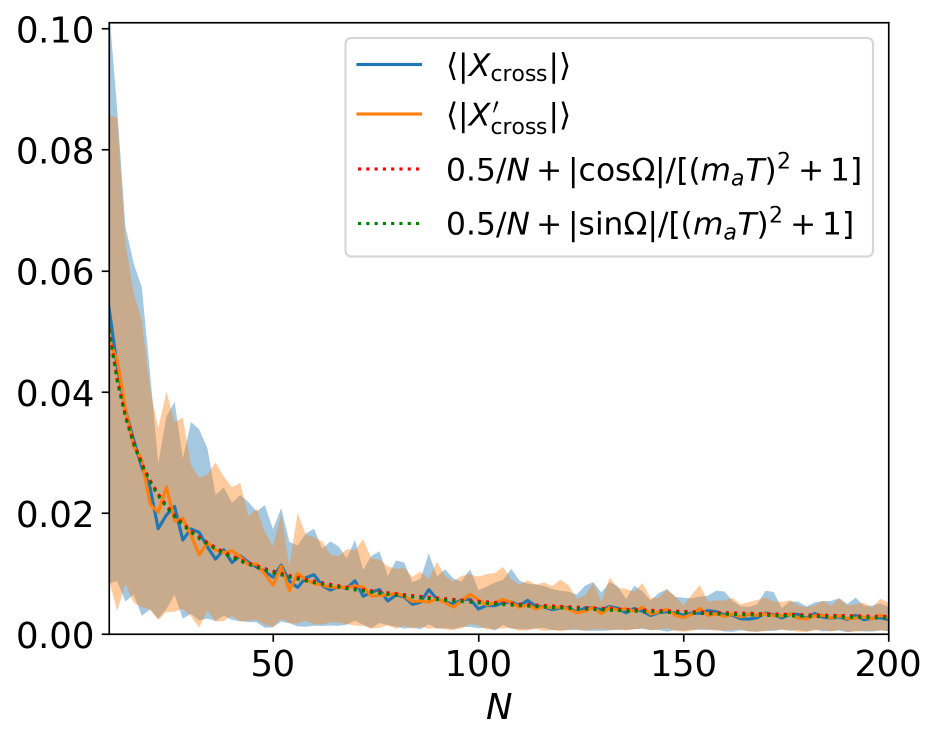}}
{ \includegraphics[width=6.5cm]{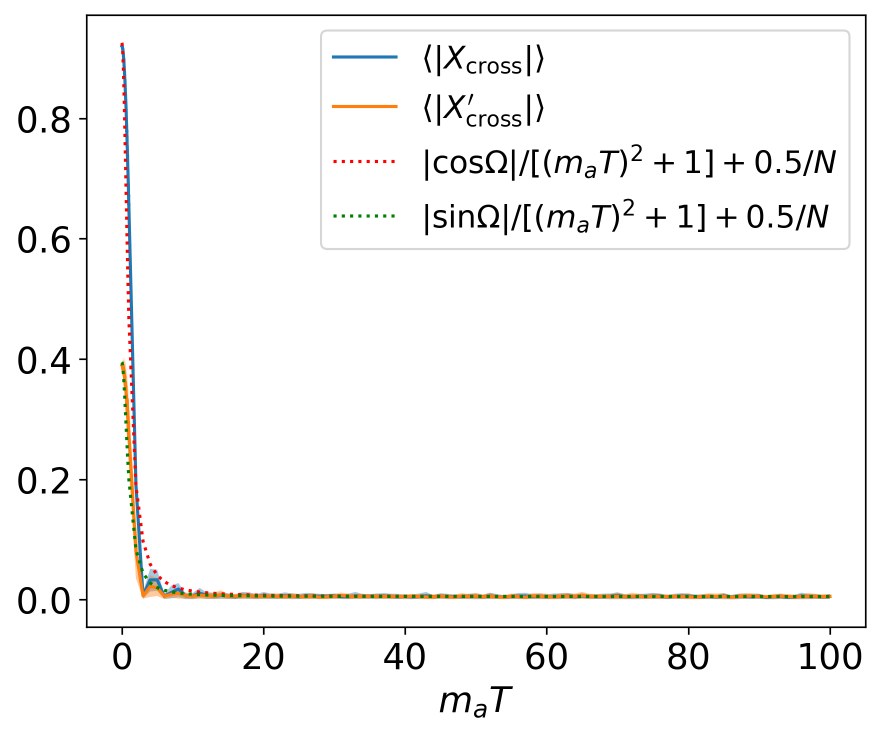}}
\caption{\label{fig:trpq} 
Dependence of  $\langle |X_{\rm cross}|\rangle $ and $\langle |X'_{\rm cross}|\rangle$ on $N$ for $m_a T=48$ (top) and on $m_a T$ for $N=100$ (bottom). Here $\langle |X_{\rm cross}|\rangle $ and $\langle |X'_{\rm cross}|\rangle$ are taken as a mean over 100 random samplings of $\{t_{p,n}; n = 1, ......, N_p\}$ and $\{t_{q,m}; m = 1, ......, N_q\}$, with the assumption of $N_p=N_q = N$, $T_p=T_q = T$ and $\Omega = 0.4$. 
The shaded bands denote their variance at 68\% C.L. The red and green dotted lines in each panel represent a simple fit to the trend of $\langle |X_{\rm cross}|\rangle $ and $\langle |X'_{\rm cross}|\rangle$, respectively.}
\end{figure}

To have some basic idea on the properties of $X_{\rm cross}$ and $X'_{\rm cross}$, let us consider a simplified case with $N_p=N_q = N$ and $T_p=T_q = T$, and randomly sample the data points. 
As shown in Fig.~\ref{fig:trpq}, $\langle |X_{\rm cross}|\rangle $ and $\langle |X'_{\rm cross}|\rangle$ in this case drop quickly to zero as $N$ and $m_a T$ increase, also with a variance of $\lesssim 0.1$. 
This trend can be well described as 
\begin{eqnarray}
\langle |X_{\rm cross}|\rangle \rightarrow \frac{0.5}{N}+\frac{|\cos \Omega|}{(m_a T)^2 +1}    \, ,  \nonumber \\
\langle |X'_{\rm cross}|\rangle \rightarrow  \frac{0.5}{N}+\frac{|\sin \Omega|}{(m_a T)^2 +1}  \, ,
\end{eqnarray}
with an injected $\Omega$ value. As $N \gg 1$, we have $X_{\rm cross}\lesssim 1$ and $X’_{\rm cross}\lesssim 1$ within the statistical range of 68\%.
Notably, $\textrm{Tr}\left(\mathbf{\Sigma}^{(s)}_{pq}\mathbf{\Sigma}^{(s)}_{qp}\right)$ is dominated by the $\textrm{Tr}\left(\hat{\mathbf{\Sigma}}^{(s)}_{pq}\hat{\mathbf{\Sigma}}^{(s)}_{qp}\right)$ and $\textrm{Tr}\left(\hat{\mathbf{\Sigma}}'^{(s)}_{pq}\hat{\mathbf{\Sigma}}'^{(s)}_{qp}\right)$ terms, while  
the $\textrm{Tr}\left(\hat{\mathbf{\Sigma}}^{(s)}_{pq}\hat{\mathbf{\Sigma}}'^{(s)}_{qp}\right)$ term becomes relevant only for $m_a T \lesssim 1$ (where $\langle |X'_{\rm cross}|\rangle \rightarrow  |\sin \Omega|$).

Generalizing this discussion, for sufficiently large $N_{p,q}$, we have 
the approximate relation in Eq.~(\ref{eq:trsigmapq}), where 
\begin{eqnarray}
&& f(y_{ep},y_{eq}) \ = \  \\
&& \ -2\rho_e^{3/2}\Big[\sqrt{\rho_p}\cos(m_aL_p)\sinc y_{ep}+(p\to q)\Big] \nonumber \\ 
&& \ +\rho_e\Big[\rho_p\sinc^2y_{ep}+ \sqrt{\rho_p\rho_q}\cos(m_a(L_p+L_q)) \nonumber\\
 &&\quad \quad  \times \sinc y_{ep}\sinc y_{eq} + (p\to q)\Big] \nonumber   \\
&& \ -2\rho_e^{1/2}\Big[\rho_p\sqrt{\rho_q}\cos(m_a L_q)\sinc y_{pq}\sinc y_{ep}+(p\to q)\Big]  \nonumber 
\end{eqnarray}
is a function of $y_{ep}$ and $y_{eq}$, and is suppressed for remote pulsars with $y_{ep}, y_{eq} \gg 1$. Note, we have dropped the $\textrm{Tr}\left(\hat{\mathbf{\Sigma}}^{(s)}_{pq}\hat{\mathbf{\Sigma}}'^{(s)}_{qp}\right)$ term in Eq.~(\ref{eq:trsigmapq}). A full treatment does not change the qualitative discussions below this equation.

\bibliography{References}

\begin{thebibliography}{56}
\bibitem[{foo()}]{footnote2}
\bibinfo{note}{We will generate the exclusion limits on the ALDM through its
  overall contribution to the test statistics, where the ALDM stochastic nature
  matters. For discovery, however, estimating the characteristic oscillation
  frequency of signals when $m_a\ll 1/T_p$ or $m_a \gg N_p/T_p$ is highly
  non-trivial}.
\end{thebibliography}


\end{document}